\documentclass[sigconf]{acmart}

\AtBeginDocument{%
  }

\setcopyright{acmlicensed}
\copyrightyear{2018}
\acmYear{2018}
\acmDOI{XXXXXXX.XXXXXXX}

\acmConference[Conference acronym 'XX]{Make sure to enter the correct
  conference title from your rights confirmation emai}{June 03--05,
  2018}{Woodstock, NY}
\acmISBN{978-1-4503-XXXX-X/18/06}

\usepackage{algorithm}
\usepackage{algpseudocode}
\usepackage{amsmath}
\usepackage{enumitem}

\begin{document}

\title{Dreaming User Multimodal Representation Guided by The Platonic Representation Hypothesis for Micro-Video Recommendation}

\author{Chengzhi Lin}
\authornote{Both authors contributed equally to this research.}
\affiliation{%
  \institution{Kuaishou Technology}
  \city{Beijing}
  \country{China}
}
\email{1132559107@qq.com}

\author{Hezheng Lin}
\authornotemark[1]
\affiliation{%
  \institution{Kuaishou Technology}
  \city{Beijing}
  \country{China}
}
\email{linhezheng@kuaishou.com}


\author{Shuchang Liu}
\affiliation{%
  \institution{Kuaishou Technology}
  \city{Beijing}
  \country{China}
}
\email{liushuchang@kuaishou.com}

\author{Cangguang Ruan}
\affiliation{%
  \institution{Kuaishou Technology}
  \city{Beijing}
  \country{China}
}
\email{ruancanguang@kuaishou.com}

\author{LingJing Xu}
\affiliation{%
  \institution{Kuaishou Technology}
  \city{Beijing}
  \country{China}
}
\email{xulingjing@kuaishou.com}

\author{Dezhao Yang}
\affiliation{%
  \institution{Kuaishou Technology}
  \city{Beijing}
  \country{China}
}
\email{yangdezhao@kuaishou.com}

\author{Chuyuan Wang}
\affiliation{%
  \institution{Kuaishou Technology}
  \city{Beijing}
  \country{China}
}
\email{wangchuyuan@kuaishou.com}


\author{Yongqi Liu}
\affiliation{%
  \institution{Kuaishou Technology}
  \city{Beijing}
  \country{China}
}
\email{liuyongqi@kuaishou.com}

\renewcommand{\shortauthors}{Trovato et al.}

\begin{abstract}
The proliferation of online micro-video platforms has underscored the necessity for advanced recommender systems to mitigate information overload and deliver tailored content. Despite advancements, accurately and promptly capturing dynamic user interests remains a formidable challenge. Inspired by the Platonic Representation Hypothesis, which posits that different data modalities converge towards a shared statistical model of reality, we introduce DreamUMM (Dreaming User Multi-Modal Representation), a novel approach leveraging user historical behaviors to create real-time user representation in a  multimoda space. DreamUMM employs a closed-form solution correlating user video preferences with multimodal similarity, hypothesizing that user interests can be effectively represented in a unified multimodal space. Additionally, we propose Candidate-DreamUMM for scenarios lacking recent user behavior data, inferring interests from candidate videos alone. Extensive online A/B tests demonstrate significant improvements in user engagement metrics, including active days and play count. The successful deployment of DreamUMM in two micro-video platforms with hundreds of millions of daily active users, illustrates its practical efficacy and scalability in personalized micro-video content delivery. Our work contributes to the ongoing exploration of representational convergence by providing empirical evidence supporting the potential for user interest representations to reside in a multimodal space.

\end{abstract}




\maketitle

\section{Introduction}

The exponential growth of micro-video platforms like TikTok, Instagram Reels, and Kuaishou has revolutionized content consumption patterns, presenting both opportunities and challenges for recommender systems. While these platforms offer unprecedented access to diverse, short-form content, they also demand sophisticated algorithms capable of capturing users' rapidly evolving interests in real-time. The ephemeral nature of micro-video consumption, characterized by users watching numerous videos in quick succession, poses a unique challenge: how to accurately model and predict user preferences in an environment where interests can shift dramatically within a single session.

Traditional approaches to user interest modeling have primarily focused on developing complex neural network architectures or refining optimization objectives to better integrate user feedback and content features\cite{micro_MARNET,micro_thacil, D2Q, D2Co, WTG, micro_fine}. However, these methods often fall short in explicitly representing user interests in a unified multimodal space, limiting their ability to capture the nuanced interplay between different content modalities that shape user preferences.

\begin{figure}[t]
  \centering
   \includegraphics[width=1.0\linewidth]{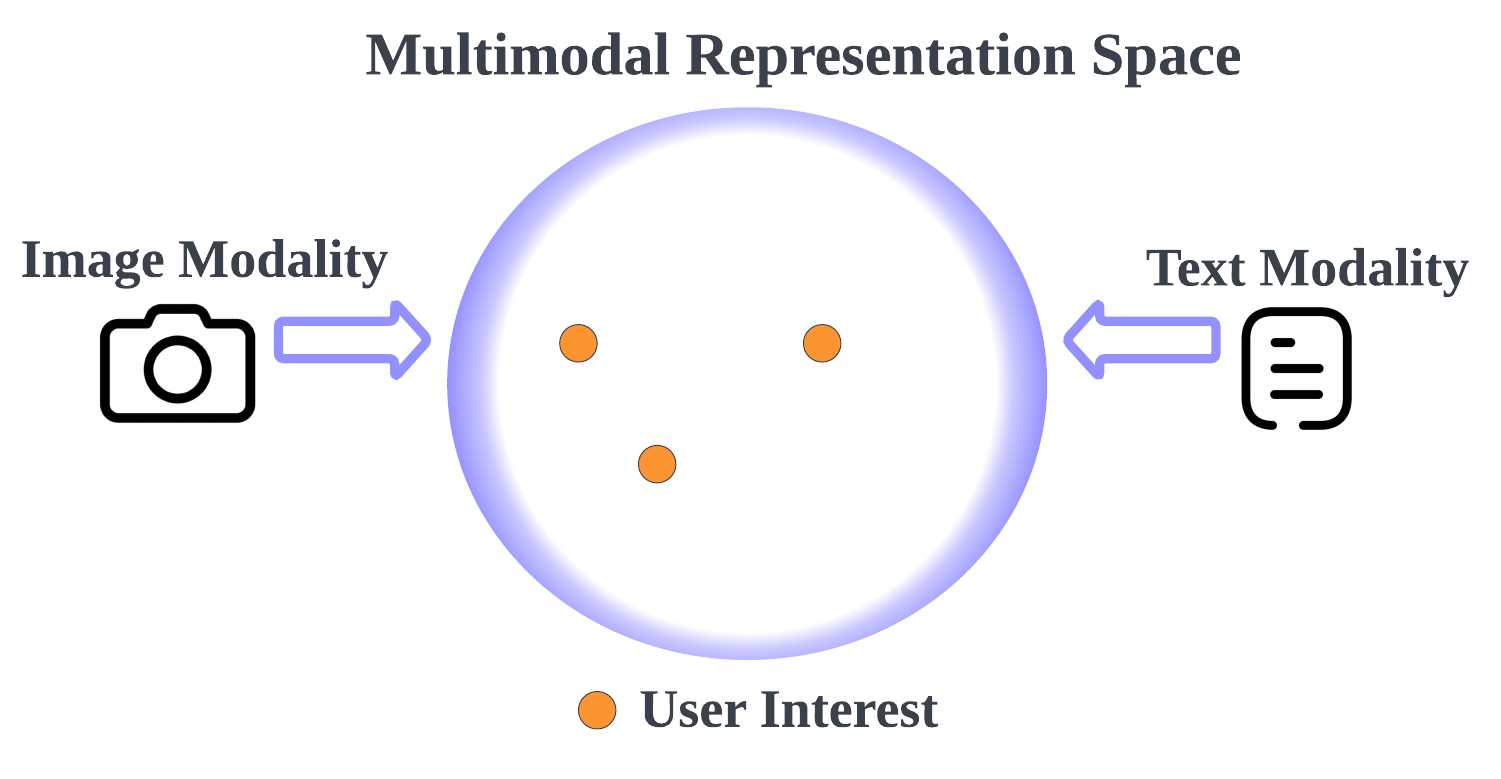}
  \caption{We hypothesize that user interests can be represented in a multimodal space, into which different data modalities (e.g., images and text) are projected.}
  \label{fig:motivation}
\end{figure}

Inspired by the Platonic Representation Hypothesis \cite{platonic}, which posits that representations of different data modalities are converging towards a shared statistical model of reality, we propose a novel approach to user interest modeling in the micro-video domain. As shown in Figure \ref{fig:motivation}, we hypothesize that an effective user interest representation can reside in the same multimodal space as the content itself, potentially offering a more holistic and accurate capture of user preferences across different modalities.
Building on this hypothesis, we introduce DreamUMM (Dreaming User Multi-Modal Representation), a novel framework for real-time user interest modeling in micro-video recommendation. DreamUMM leverages users' historical interactions to generate multimodal representations that reflect their dynamic interests, guided by the principle that a user's affinity towards a video should positively correlate with their similarity in the multimodal space. To address scenarios where recent user behavior data is unavailable, such as when users reopen the app after extended intervals, we propose Candidate-DreamUMM, a variant designed to infer user interests solely based on candidate videos.

Central to our approach is a novel multimodal representation learning framework that leverages large language models and knowledge distillation to create rich, informative video representations. This framework forms the foundation of both DreamUMM and Candidate-DreamUMM, enabling the creation of high-quality multimodal embeddings that capture the complex interplay between visual, auditory, and textual elements in micro-videos.
Extensive online A/B tests demonstrate the effectiveness of our proposed methods, showing significant improvements in key user engagement metrics, including  active days and play count. The successful deployment of DreamUMM and Candidate-DreamUMM in two major micro-video platforms, serving hundreds of millions of users, further validates the practical utility and scalability of our approach in real-world scenarios.
The main contributions of our work are as follows:
\begin{itemize}

    \item We propose DreamUMM, a novel user representation learning framework that models user interests in a multimodal space, drawing inspiration from the Platonic Representation Hypothesis.
    \item We introduce Candidate-DreamUMM, an extension specifically designed to address the cold-start problem and capture users' current interests by focusing on candidate videos.
    \item We develop a multimodal representation learning framework that leverages large language models and knowledge distillation to create high-quality video embeddings.
    \item We conduct extensive online experiments and real-world deployments to demonstrate the effectiveness and practical impact of both DreamUMM and Candidate-DreamUMM.
    \item Our work contributes to the ongoing exploration of representational convergence by providing empirical evidence supporting the potential for user interest representations to reside in a multimodal space.
\end{itemize}
By bridging the gap between theoretical insights from the Platonic Representation Hypothesis and practical recommender system design, our work not only advances the state-of-the-art in micro-video recommendation but also opens new avenues for research in multimodal user modeling and content understanding. The success of our approach suggests that future recommender systems may benefit from explicitly modeling user interests in unified multimodal spaces, potentially leading to more accurate, versatile, and interpretable recommendations across various domains.

\section{Method}

\subsection{Problem Formulation}
In the domain of micro-video recommendation, accurately capturing users' dynamic interests in real-time is crucial. Let $\mathcal{U}$ and $\mathcal{V}$ denote the sets of users and micro-videos, respectively. For each user $u \in \mathcal{U}$, we have access to their historical interaction sequence $\mathcal{I}_u = \{(v_j, r_j)\}_{j=1}^N$, where $v_j \in \mathcal{V}$ represents the $j$-th micro-video watched by user $u$, and $r_j$ indicates the corresponding interaction strength (e.g., watch time, likes, comments).

Our goal is to learn a function $f: \mathcal{U} \rightarrow \mathbb{R}^d$ that maps each user to a $d$-dimensional representation space, capturing their real-time interests based on their historical interactions. This representation should effectively model the rapid shifts in user preferences characteristic of micro-video consumption.
Existing methods for user interest modeling, such as recurrent neural networks and self-attention mechanisms\cite{SIM}, often lack an explicit mechanism to map user interests into a multimodal representation space. This limits their ability to capture users' preferences across modalities. Our approach aims to address this limitation by leveraging insights from the Platonic Representation Hypothesis\cite{platonic}.

\subsection{The Platonic Representation Hypothesis for User Interests}

\begin{figure}[t]
  \centering
   \includegraphics[width=0.9\linewidth]{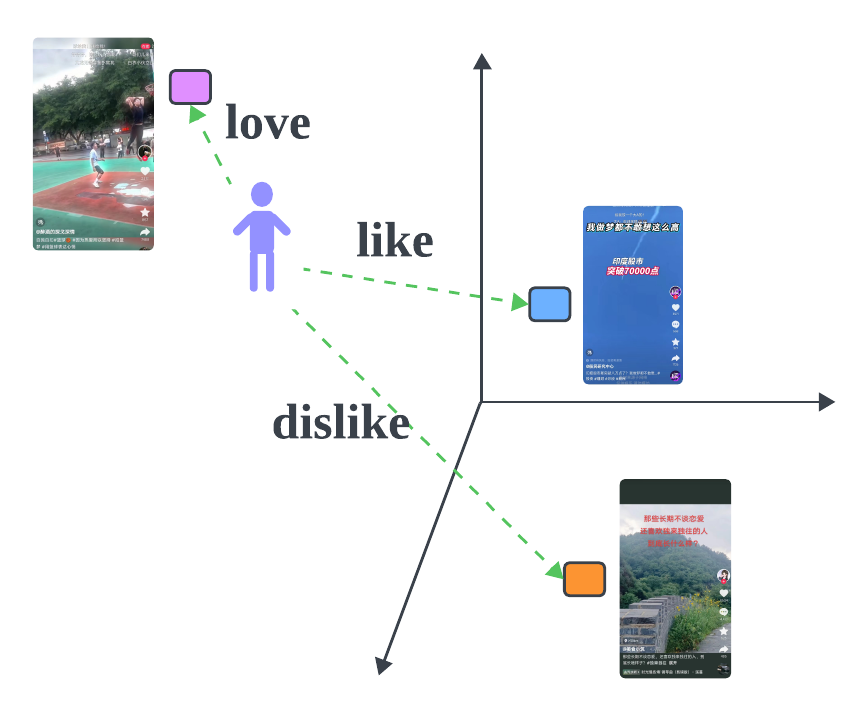}
  \caption{DreamUMM constructs the user's multimodal representation based on the user's liking for micro videos.}
  \label{fig:dreamumm}
\end{figure}

Recently, the Platonic Representation Hypothesis \cite{platonic} proposed that different data modalities are converging towards a unified representation that reflects objective reality. Inspired by this concept, we hypothesize that users' interest representations may reside in a multimodal space that is shared with the space of video content. This hypothesis is based on two key assumptions:
\begin{enumerate}
\item User interests are grounded in their perception and understanding of the real world, which is shaped by their interactions with content across different modalities.
\item If representations of different data modalities are indeed converging towards a unified multimodal space that effectively captures the real world, it is plausible that user interests can also be represented in this space.
\end{enumerate}
Formally, we posit that there exists a latent multimodal space $\mathcal{Z}$ that encapsulates both video content and user interests. In this space, we aim to learn a user representation $\mu_u \in \mathcal{Z}$ for each user $u$, such that:
\begin{equation}
\mu_u = f(\mathcal{I}_u)
\end{equation}
where $f$ is a function that maps the user's interaction history to the multimodal space $\mathcal{Z}$.

Building on this hypothesis, we propose DreamUMM (Dreaming User Multi-Modal Representation), a novel approach for real-time user interest modeling in the context of micro-video recommendation.

\subsection{DreamUMM: Dreaming User Multi-Modal Representation}
DreamUMM leverages users' historical interactions to generate multimodal representations that reflect their dynamic interests. The key idea is to construct a user representation that is close to the representations of videos they prefer in the multimodal space.
\subsubsection {User Multimodal Representation}
Given a user's historical interaction sequence $\mathcal{I}_u = \{(u, v_j, r_j)\}_{j=1}^M$, we aim to produce a multimodal representation $\mu_{hist}$ for the user in the shared multimodal space $\mathcal{Z}$. Let $\mathbf{x}_j \in \mathcal{Z}$ be the multimodal representation of video $v_j$, derived from pre-trained multimodal models.
As shown in Figure \ref{fig:dreamumm}, we propose the following optimization criterion:
\begin{equation}
\mu_{hist} = \text{argmax}_{\mu, \|\mu\|=1} \sum_{j=1}^M a_j \langle \mathbf x_{j}, \mu\rangle.
\end{equation}
where $a_j$ represents the user's preference for video $v_j$, and $\langle \cdot, \cdot \rangle$ denotes the inner product.
This formulation has a closed-form solution:
\begin{equation}
\begin{aligned}
    \mu_{hist} 
        &=\frac {\sum_{j=1}^M a_j \mathbf x_{j}} {\|\sum_{j=1}^M a_j \mathbf x_{j}\|}
\end{aligned}
\end{equation}
\subsubsection{User Preference Scoring}
Inspired by D2Q \cite{D2Q}, we define the user's preference score $a_j$ for video $v_j$ as:
\begin{equation} \label{eq:long_view}
a_j = \frac{1}{1 + \exp(-\alpha(w_j - t_j))}
\end{equation}
where $w_j$ is the user's watched time on video $v_j$, $t_j$ is a long-view threshold, and $\alpha$ controls the sensitivity between preference and watched time. This soft thresholding function accounts for the noise inherent in online behaviors.
\subsubsection{Theoretical Justification}
The DreamUMM approach aligns with the Platonic Representation Hypothesis in several ways:
\begin{itemize}
\item It explicitly represents user interests in the same multimodal space as video content, reflecting the hypothesis of a shared underlying reality.
\item The use of pre-trained multimodal models to obtain video representations $\mathbf{x}_j$ leverages the convergence of different modalities towards a unified representation.
\item The optimization criterion encourages the user representation to be similar to preferred video representations, potentially capturing the user's understanding of the "real world" as reflected in their video preferences.
\end{itemize}

\begin{algorithm}
\caption{Online Recommendation Process with DreamUMM and Candidate-DreamUMM} \label{alg:all}
\begin{algorithmic}[1]
\Function{ProcessUserRequest}{$u, I_u, V, \text{useCandidate}$} \Comment{User $u$, historical interactions $I_u$, candidate videos $V$, boolean flag useCandidate}
    \State \textbf{Output:} Ranked list of recommended videos
    
    \If{useCandidate}
        \State $\mu \gets \text{CandidateDreamUMM}(V, I_u)$
    \Else
        \State $\mu \gets \text{DreamUMM}(I_u)$
    \EndIf
    \For{each video $v_i$ in $V$}
        \State $s_i \gets \langle \text{MultimodalRepresentation}(v_i), \mu \rangle$
    \EndFor
    \State Sort $V$ based on similarity scores $s_i$ and other scores
    \State \Return Top-k videos from sorted $V$
\EndFunction

\Function{DreamUMM}{$I_u$} \Comment{Historical interactions $I_u = \{(v_j, r_j)\}$}
    \State \textbf{Output:} User representation $\mu_{hist}$
    \For{each interaction $(v_j, r_j)$ in $I_u$}
        \State Compute $a_j$ using Eq. \ref{eq:long_view}
        \State $x_j \gets \text{MultimodalRepresentation}(v_j)$
    \EndFor
    \State $\mu_{hist} \gets \frac{\sum_{j=1}^M a_j x_j}{\|\sum_{j=1}^M a_j x_j\|}$
    \State \Return $\mu_{hist}$
\EndFunction

\Function{CandidateDreamUMM}{$V, I_u$} \Comment{Candidate videos $V$, historical interactions $I_u$}
    \State \textbf{Output:} User representation $\mu_{candidate}$
    \For{each video $v_i$ in $V$}
        \State $a_i \gets f_{seq}(I_u, v_i)$ \Comment{Predict preference score}
        \State $x_i \gets \text{MultimodalRepresentation}(v_i)$
    \EndFor
    \State $\mu_{candidate} \gets \frac{\sum_{i=1}^N a_i x_i}{\|\sum_{i=1}^N a_i x_i\|}$
    \State \Return $\mu_{candidate}$
\EndFunction
\end{algorithmic}
\end{algorithm}
\subsection{Candidate-DreamUMM: Addressing Cold-Start Scenarios}
While DreamUMM effectively captures user interests based on historical interactions, it may face challenges in scenarios where recent user behavior data is unavailable, such as when a user reopens the app after an extended period. To address this issue, we propose Candidate-DreamUMM, a variant designed to infer user interests solely based on the current context, i.e., the candidate videos.
\subsubsection{Motivation}
The motivation behind Candidate-DreamUMM is twofold:
\begin{itemize}
\item It tackles the cold-start problem when recent user behavior data is unavailable.
\item It captures users' current interests more accurately by focusing on the candidate videos, which reflect the present context and are more likely to align with users' immediate preferences.
\end{itemize}
\subsubsection{Formulation}
For a given set of candidate videos $\{v_i\}_{i=1}^N$, Candidate-DreamUMM constructs a user representation as follows:
\begin{equation}
\mu_{candidate} = \text{argmax}_{\mu, |\mu|=1} \sum_{i=1}^N a_i \langle \mathbf{x}_i, \mu \rangle
\end{equation}
The closed-form solution is:
\begin{equation}
\mu_{candidate} = \frac {\sum_{i=1}^N a_i \mathbf x_{i}} {\|\sum_{i=1}^N a_i \mathbf x_{i}\|}
\end{equation}
where $\mathbf{x}_i$ is the multimodal representation of candidate video $v_i$, and $a_i$ is the predicted preference score for the candidate video.
\subsubsection{Preference Score Prediction}
In Candidate-DreamUMM, the preference score $a_i$ is predicted by an online sequence model:
\begin{equation}
a_i = f_{seq}(\mathcal{I}_u, v_i)
\end{equation}
where $f_{seq}$ is a sequence model that takes the user's historical interaction list $\mathcal{I}_u$ and the candidate video $v_i$ as input, and outputs the predicted long-view probability.
\subsubsection{Theoretical Connection}
Candidate-DreamUMM maintains the core idea of the Platonic Representation Hypothesis by:
\begin{itemize}
\item Representing user interests in the same multimodal space as video content.
\item Leveraging the predicted preferences on candidate videos to infer the user's current position in the multimodal space.
\item Adapting to the user's evolving interests by focusing on the current context, aligning with the dynamic nature of the "real world" representation.
\end{itemize}


\subsection{Multimodal Representation Learning}
\begin{figure*}[t]
  \centering
   \hfill
   \includegraphics[width=0.9\linewidth]{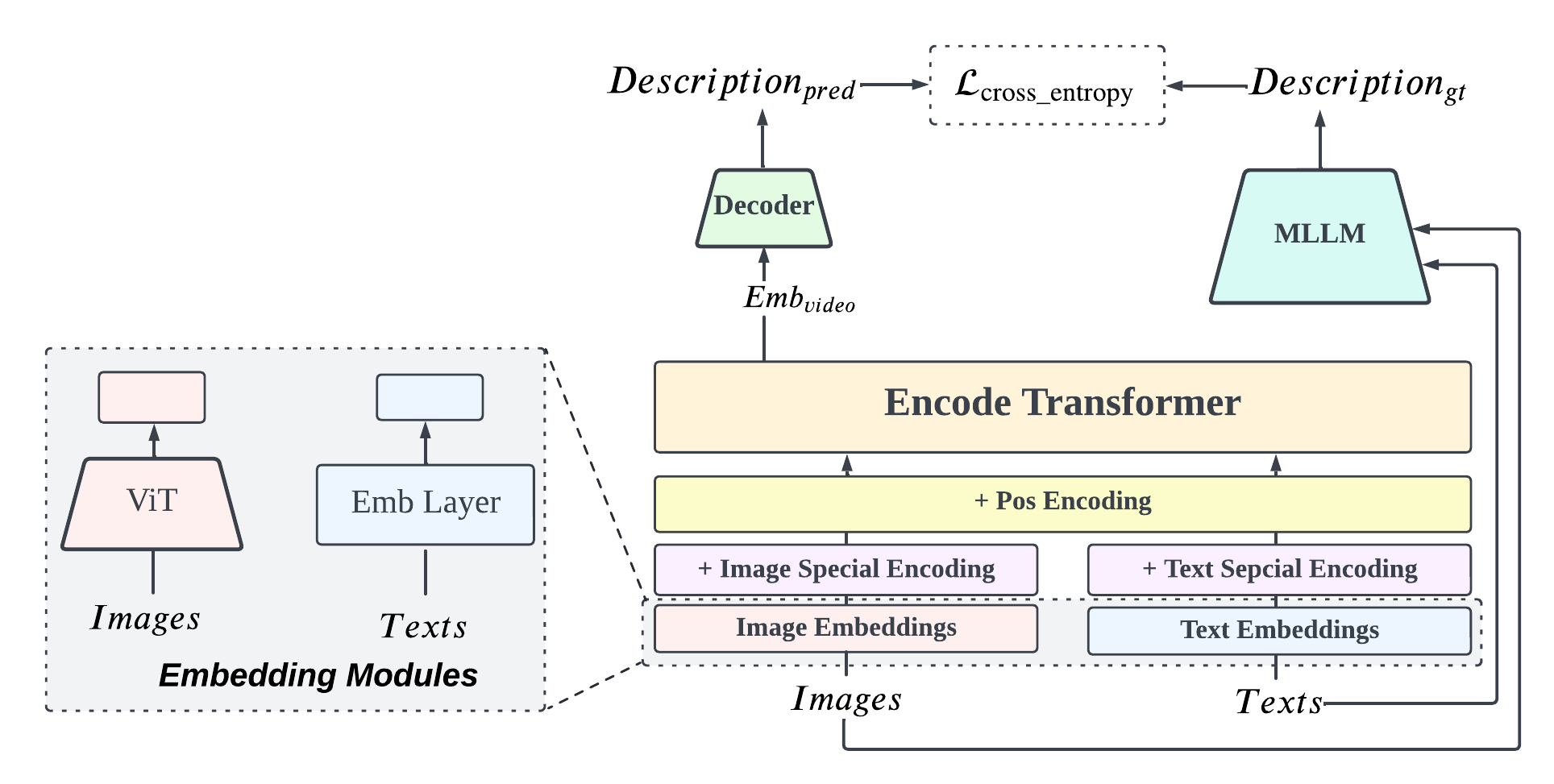}
  \vspace{-0.5cm}   
  \caption{Multimodal representation learning framework.}
  \hfill
  \label{fig:representation}
  \vspace{-0.5cm}
\end{figure*}
A critical component of our approach is the learning of high-quality multimodal representations for videos. These representations form the foundation of both DreamUMM and Candidate-DreamUMM. We propose a novel framework that leverages large language models and knowledge distillation to create rich, informative video representations.
\subsubsection{Motivation}
Videos are inherently multimodal, containing visual, auditory, and textual information. Capturing the nuances of these different modalities and their interactions is crucial for effective recommendation. While large multimodal models have shown impressive capabilities in understanding such complex data, their computational requirements make them impractical for real-time recommendation systems. Our goal is to distill the knowledge from these large models into a more efficient representation.
\subsubsection{Framework Overview}
Our multimodal representation learning framework consists of the following key components:
\begin{enumerate}
\item A Multimodal Large Language Model (MLLM) for generating comprehensive video descriptions.
\item An encoder-decoder architecture for learning compact video representations.
\item A knowledge distillation process to transfer information from the MLLM to our efficient model.
\end{enumerate}
\subsubsection{Detailed Methodology}
\paragraph{MLLM-based Video Description}
We utilize a pre-trained MLLM to generate detailed descriptions of videos, including themes, characters, scenes, and other relevant information. These descriptions serve as a rich supervisory signal for our representation learning model.
\paragraph{Encoder-Decoder Architecture}
Our model consists of:
\begin{itemize}
\item An encoder that processes multimodal inputs (e.g., video frames, audio features, metadata).
\item A fully connected layer that condenses the multimodal tokens into a single video token representation.
\item A decoder that generates the comprehensive description produced by the MLLM, using the video token as key and value inputs.
\end{itemize}
Formally, let $E(\cdot)$ be the encoder, $D(\cdot)$ the decoder, and $F(\cdot)$ the fully connected layer. The video representation $\mathbf{x}$ is computed as:
\begin{equation}
\mathbf{x} = F(E(v))
\end{equation}
where $v$ represents the multimodal inputs of the video.
\paragraph{Training Objective}
We train our model using a cross-entropy loss between the generated description and the MLLM-produced description:
\begin{equation}
\mathcal{L} = -\sum_{i} y_i \log(D(\mathbf{x})_j)
\end{equation}
where $y$ is the one-hot encoded MLLM description, and $D(\mathbf{x})_j$ is the model's predicted probability for token $j$.
\subsubsection{Theoretical Justification}
This approach aligns with the Platonic Representation Hypothesis in several ways:
\begin{itemize}
\item It leverages the MLLM's ability to generate unified representations across modalities.
\item The distillation process transfers this unified understanding to our more efficient model.
\item The resulting video representations capture rich, multimodal information about the video content, potentially approaching the "ideal" representation of reality posited by the hypothesis.
\end{itemize}

\subsection{Online Application}

Algorithm \ref{alg:all} presents the core components of our online recommendation process, integrating DreamUMM and Candidate-DreamUMM into a flexible, real-time recommendation workflow.

The main function, ProcessUserRequest (lines 1-13), handles each user request for recommendations. It takes four inputs: the user $u$, their historical interactions $I_u$, a set of candidate videos $V$, and a boolean flag useCandidate. This flag allows the system to dynamically choose between DreamUMM and Candidate-DreamUMM based on various factors such as the recency and sufficiency of the user's historical interactions, or other contextual information.

In our online system, the process flows as follows:
\begin{enumerate}
    \item  When a user requests recommendations, ProcessUserRequest is called with the appropriate parameters, including the useCandidate flag.

    \item  Based on the useCandidate flag, either DreamUMM or Candidate-DreamUMM is used to generate the user's representation.

    \item The function then computes similarity scores between the user representation and each candidate video using their multimodal representations.

    \item Finally, it ranks the candidate videos based on these similarity scores, potentially combining them with other relevance signals, and returns the top-k recommendations.

\end{enumerate}
This approach allows us to efficiently generate personalized recommendations in real-time, adapting to both the user's historical preferences and the current context of available videos. By providing the flexibility to choose between DreamUMM and Candidate-DreamUMM at runtime, our system can handle various scenarios of user data availability and recommendation contexts, ensuring robust and personalized recommendations for all users.

The integration of multimodal representations throughout this process, from user modeling to video similarity computation, enables our system to capture rich, cross-modal information about both users and content. This aligns with our hypothesis that user interests can be effectively represented in a unified multimodal space, potentially leading to more accurate and diverse recommendations.



\begin{table*}
    \centering
    \begin{tabular}{cccccc}
        \hline
        Method & \multicolumn{2}{c}{Platform A} & \multicolumn{2}{c}{Platform B} \\ 
        & Active Days & Play Count  & Active Days & Play Count  \\ \hline
        DreamUMM & 0.003\% & \textbf{0.273\%} &  0.000\% & \textbf{0.287\%}  \\ \hline
        Candidate-DreamUMM & \textbf{0.037\%} & \textbf{0.867\%}  & \textbf{0.050\%} & \textbf{0.318\%}  \\ \hline
    \end{tabular}
    \caption{Results of online A/B experiments, measured by Active Days and Play Count. Each row indicates the relative improvement with our method over the online baseline, which already includes the SIM (Search-based user Interest Model) model \cite{SIM}. Statistically significant improvement is marked with bold font in the table (p-value < 5\%).}
    \label{tab:online}
\end{table*}

\begin{table}[t]
    \centering
    \begin{tabular}{ccccccc}
        \hline
        Method & HitRate@100 & HitRate@200 & \\ \hline
        Representation/wo.MLLM & 0.730 & 0.678 & \\ \hline
        Representation/w.MLLM(ours) & \textbf{0.742} & \textbf{0.688} & \\ \hline
    \end{tabular}
    \caption{The results of our offline video retrieval benchmark demonstrate that our new representation, enhanced by the distillation process of a MLLM, has achieved significant and consistent improvements in precision. Specifically, 'HitRate@100' indicates the mean precision for the top 50 recall videos across a set of 100 query videos, while 'HitRate@200' applies this metric to an expanded set of 200 query videos, underscoring the robustness and reliability of our approach in enhancing retrieval accuracy.}
    \label{tab:offline}
\end{table}

\section{Experiments and Results}
In this section, we present a comprehensive evaluation of DreamUMM and Candidate-DreamUMM through both online and offline experiments. The experiments are designed to answer the following research questions:
\begin{enumerate}[leftmargin=*, align=left, labelsep=0pt]
\item RQ1: How do DreamUMM and Candidate-DreamUMM perform in terms of improving user engagement in real-world micro-video platforms?
\item RQ2: How effective are DreamUMM and Candidate-DreamUMM in enhancing recommendation diversity and expanding users' interest range?
\item RQ3: How well does our multimodal representation learning framework capture video semantics and support accurate retrieval?
\end{enumerate}

%
\subsection{Experimental Setup}
\textbf{Online Experiments}: We conducted online A/B tests on two popular micro-video platforms, denoted as Platform A and Platform B. Each platform has hundreds of millions of daily active users (DAU). For each platform, we randomly split users into control and treatment groups, with at least 10\% of the total DAU in each group. 

We employed several metrics to evaluate the online performance:
\begin{itemize}

    \item Play Count: The average number of micro-videos played per user during the experiment period.
    \item Active Days: The average number of active days per user within the experiment duration. An active day is defined as a day when the user plays at least one micro-video.
    \item  Exposed Cluster: The average number of unique clusters that a user is exposed to in the recommended video list in eacy day. The clusters are generated based on video content similarity, with each cluster representing a group of semantically similar videos. A higher Exposed Cluster Count indicates a more diverse recommendation list covering a wider range of user interests.
    \item  Surprise Cluster: The proportion of recommended micro-videos that are dissimilar to users' historical preferences yet receive high positive feedback.
\end{itemize}

\vspace{0.1cm}
\noindent\textbf{Offline experiments:} To validate the quality of our learned video representations, we constructed a video retrieval dataset containing about 40,000 micro videos annotated by human experts. We utilized HitRate as our primary evaluation metric, defined as:
\begin{equation}
    HitRate = \frac{1}{N}\sum_{i=1}^{N}\frac{R_i}{L_i}
\end{equation}

where $N$ denotes the number of query videos, $R_i$ represents the number of correctly retrieved relevant videos for the $i$-th query video, and $L_i$ signifies the total number of relevant videos for the $i$-th query video. We specifically employ HitRate@100 and HitRate@200 to assess model performance.

\begin{figure}[t]
  \centering
   \includegraphics[width=1.0\linewidth]{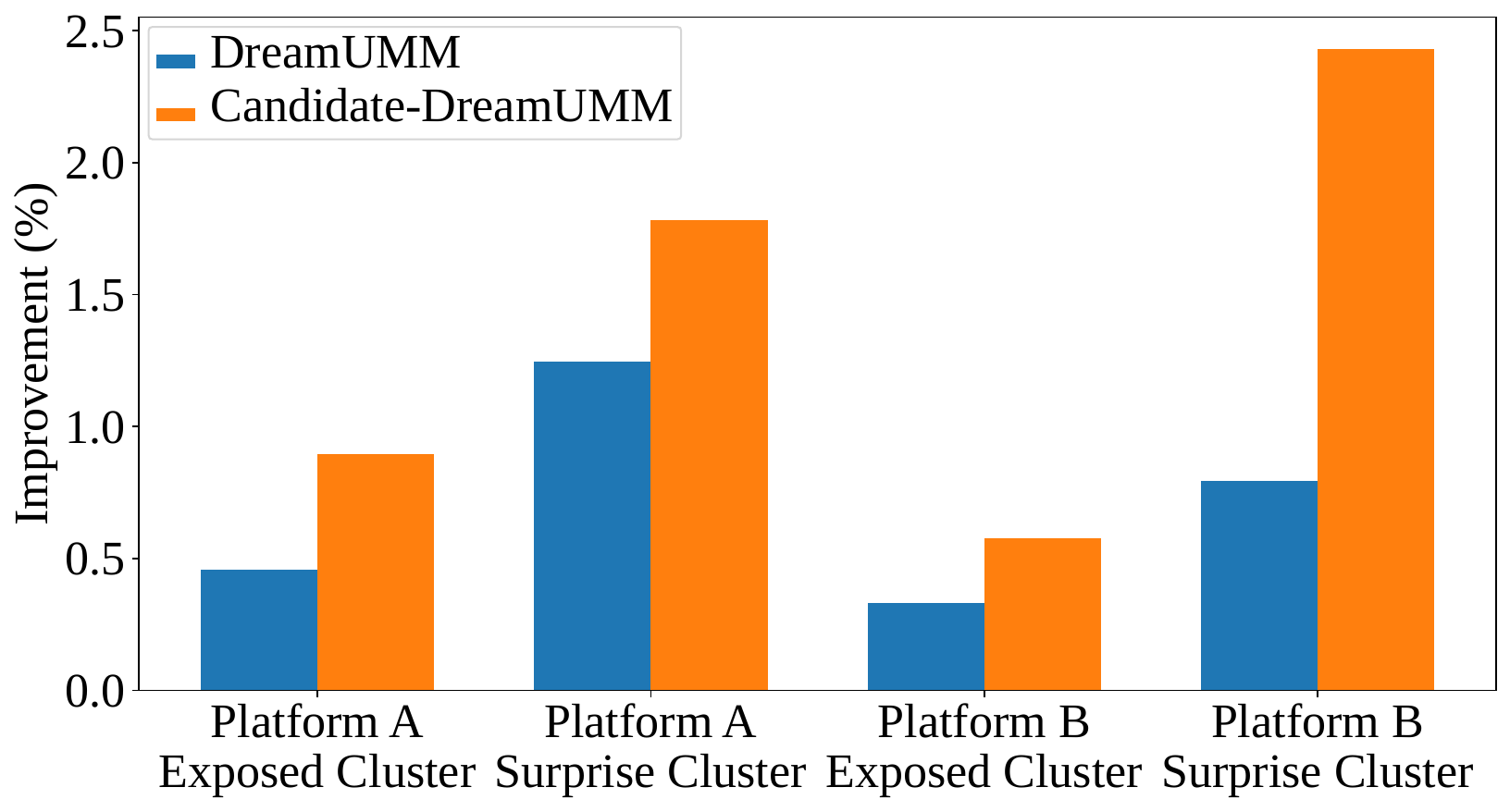}
  \caption{Diveristy Results of DreamUMM and Candidate-DreamUMM on two micro-video platforms. The bar chart illustrates the relative improvements in Exposed Cluster Count and Surprise Cluster metrics over the online method. Candidate-DreamUMM consistently outperforms DreamUMM across both platforms and metrics, with the most significant gains observed in the Surprise Cluster metric on Platform B. These results demonstrate the effectiveness of Candidate-DreamUMM in enhancing recommendation diversity and novelty by leveraging contextual information to capture users' real-time preferences.}
  \label{fig:diversity_result}
\end{figure}

\subsection{Results and Analysis}
\noindent\textbf{RQ1}: User Engagement. Table \ref{tab:online} presents the relative improvements of DreamUMM and Candidate-DreamUMM over the control group in terms of Play Count and Active Days on both platforms. We observe significant gains in both metrics, indicating the effectiveness of our methods in enhancing user engagement. Candidate-DreamUMM consistently outperforms DreamUMM, suggesting its superior ability to capture users' real-time interests by focusing on the current context. The lifts in Play Count and Active Days demonstrate that our methods can effectively encourage users to consume more videos and visit the platform more frequently.

\vspace{0.1cm}
\noindent\textbf{RQ2}: Recommendation Diversity. Figure \ref{fig:diversity_result} visualizes the improvements of DreamUMM and Candidate-DreamUMM in Exposed Cluster and Surprise Cluster metrics over the control group. Both methods achieve substantial gains in recommendation diversity, with Candidate-DreamUMM showing larger improvements. The Surprise Cluster metric sees the most impressive boost, where Candidate-DreamUMM increases the proportion of surprised recommendations by 2.429\% and 1.782\% on Platform A and Platform B, respectively. These results validate the effectiveness of our methods, especially Candidate-DreamUMM, in expanding users' interest range and enhancing recommendation diversity.

\vspace{0.1cm}
\noindent\textbf{RQ3}: Representation Quality. Table \ref{tab:offline} presents the HitRat@100 and HitRat@200 of our model with MLLM-based representation learning and a variant without MLLM pre-training. Our full model achieves a HitRate@100 of 0.742 and a HitRate@200 of 0.688, significantly outperforming the variant without MLLM pre-training. This demonstrates the effectiveness of leveraging the knowledge encoded in the MLLM to learn informative video representations that align well with human judgments of content similarity.

In summary, our experiments comprehensively demonstrate the effectiveness of DreamUMM and Candidate-DreamUMM in improving user engagement and recommendation diversity in real-world micro-video platforms. The offline evaluation further validates the quality of our learned video representations and highlights the importance of MLLM-based representation learning. The combination of online and offline results provides strong empirical evidence supporting the Platonic Representation Hypothesis, showing that our learned representations align well with the underlying content and semantics of the videos, and that modeling user interests in a unified multimodal space can lead to significant practical benefits in personalized micro-video content delivery

\section{Related Work}
\subsection{Video Recommendation}
The field of video recommendation has seen substantial advancements with the evolution of deep learning techniques and the increasing availability of user interaction data. Traditional video recommendation systems focus on collaborative filtering and content-based filtering \cite{cf_1, cf_2, cf_3, cf_4}. Collaborative filtering leverages user-item interaction matrices but often struggles with cold-start problems and sparse data scenarios. Conversely, content-based filtering \cite{content_1, content_2, content_3} utilizes video metadata and content features to recommend similar items but may not fully capture the nuanced preferences of users.

Recent approaches have integrated deep learning models to enhance the understanding of video content and user preferences. Attention mechanisms and graph neural networks (GNNs) have been employed to model the temporal dynamics of user interactions and the complex relationships between videos \cite{micro_MARNET, video_gnn}. For instance, MARNET \cite{micro_MARNET} aggregates multimodal information using a visual-centered modality grouping approach and learns dynamic label correlations through an attentive GNN.

In the video recommendation domain, where explicit feedback is sparse, some methods specifically address how to define whether a user is interested in a video through implicit feedback, utilizing techniques such as causal inference \cite{D2Co, D2Q}, fine-grained mining \cite{micro_fine}, and distribution alignment \cite{WTG, micro_len}.

Our approach diverges from the traditional focus on network design or defining interest. Inspired by the Platonic Representation Hypothesis \cite{platonic}, we concentrate on the explicit representation of user interest in a multimodal space. This method facilitates a more precise depiction of user preferences by leveraging multimodal data to construct robust user representation.

\subsection{Multimodal Recommendation}

Multimodal recommendation extends beyond traditional recommendation paradigms by incorporating diverse data modalities such as text, images, audio, and video to build a comprehensive understanding of user preferences. This approach is particularly beneficial in micro-video platforms where content is rich in multimodal features. The integration of these modalities provides a deeper semantic understanding and can significantly enhance recommendation performance.

Multimodal learning frameworks have been developed to fuse information from various sources, leveraging techniques such as graph convolution, multimodal autoencoders, attention-based fusion methods, transformer architectures and Flat Local Minima Exploration\cite{mmr_vae, mmr_att, mmr_trans, mmr_con, dragon, freedom, BM3}. For example, DRAGON \cite{dragon} utilizes user-user co-occurrence graphs in combination with item-item multimodal graphs to enhance the user-item heterogeneous graph. MG \cite{mirror} introduces a mirror-gradient method to address the training instability issues caused by multimodal input.
The challenge remains in effectively combining multimodal data to reflect real-time user preferences. 

By generating real-time user representations in a multimodal space, DreamUMM presents a practical solution for dynamic micro-video recommendation. Furthermore, our Candidate-DreamUMM variant addresses the cold start problem by inferring preferences from candidate videos alone, showcasing the flexibility and robustness of our approach in real-world applications.

\section{Conclusion}

This paper introduced DreamUMM and Candidate-DreamUMM, novel approaches for micro-video recommendation that leverage unified multimodal representations. By modeling user interests in the same multimodal space as video content, our framework addresses both dynamic preference changes and cold-start scenarios.
Through extensive online A/B tests, we demonstrated significant improvements in user engagement and recommendation novelty. The successful deployment underscores the practical efficacy and scalability of our methods.
Our work contributes empirical evidence supporting the Platonic Representation Hypothesis - the potential for user interest representations to reside in a multimodal space. This insight opens new avenues for research in multimodal user modeling and content understanding. Looking ahead, future work will focus on designing end-to-end methods to jointly learn the shared multimodal space for users and videos, potentially enhancing personalized recommendations across domains.

\bibliographystyle{ACM-Reference-Format}
\bibliography{sample-base}


\begin{thebibliography}{25}


\ifx \showCODEN    \undefined \def \showCODEN     #1{\unskip}     \fi
\ifx \showDOI      \undefined \def \showDOI       #1{#1}\fi
\ifx \showISBNx    \undefined \def \showISBNx     #1{\unskip}     \fi
\ifx \showISBNxiii \undefined \def \showISBNxiii  #1{\unskip}     \fi
\ifx \showISSN     \undefined \def \showISSN      #1{\unskip}     \fi
\ifx \showLCCN     \undefined \def \showLCCN      #1{\unskip}     \fi
\ifx \shownote     \undefined \def \shownote      #1{#1}          \fi
\ifx \showarticletitle \undefined \def \showarticletitle #1{#1}   \fi
\ifx \showURL      \undefined \def \showURL       {\relax}        \fi
\providecommand\bibfield[2]{#2}
\providecommand\bibinfo[2]{#2}
\providecommand\natexlab[1]{#1}
\providecommand\showeprint[2][]{arXiv:#2}

\bibitem[Chen et~al\mbox{.}(2018a)]%
        {micro_thacil}
\bibfield{author}{\bibinfo{person}{Xusong Chen}, \bibinfo{person}{Dong Liu}, \bibinfo{person}{Zheng{-}Jun Zha}, \bibinfo{person}{Wengang Zhou}, \bibinfo{person}{Zhiwei Xiong}, {and} \bibinfo{person}{Yan Li}.} \bibinfo{year}{2018}\natexlab{a}.
\newblock \showarticletitle{Temporal Hierarchical Attention at Category- and Item-Level for Micro-Video Click-Through Prediction}. In \bibinfo{booktitle}{\emph{2018 {ACM} Multimedia Conference on Multimedia Conference, {MM} 2018, Seoul, Republic of Korea, October 22-26, 2018}}, \bibfield{editor}{\bibinfo{person}{Susanne Boll}, \bibinfo{person}{Kyoung~Mu Lee}, \bibinfo{person}{Jiebo Luo}, \bibinfo{person}{Wenwu Zhu}, \bibinfo{person}{Hyeran Byun}, \bibinfo{person}{Chang~Wen Chen}, \bibinfo{person}{Rainer Lienhart}, {and} \bibinfo{person}{Tao Mei}} (Eds.). \bibinfo{publisher}{{ACM}}, \bibinfo{pages}{1146--1153}.
\newblock
\urldef\tempurl%
\url{https://doi.org/10.1145/3240508.3240617}
\showDOI{\tempurl}


\bibitem[Chen et~al\mbox{.}(2018b)]%
        {content_1}
\bibfield{author}{\bibinfo{person}{Xusong Chen}, \bibinfo{person}{Rui Zhao}, \bibinfo{person}{Shengjie Ma}, \bibinfo{person}{Dong Liu}, {and} \bibinfo{person}{Zheng-Jun Zha}.} \bibinfo{year}{2018}\natexlab{b}.
\newblock \showarticletitle{Content-based video relevance prediction with second-order relevance and attention modeling}. In \bibinfo{booktitle}{\emph{Proceedings of the 26th ACM international conference on Multimedia}}. \bibinfo{pages}{2018--2022}.
\newblock


\bibitem[Covington et~al\mbox{.}(2016)]%
        {content_2}
\bibfield{author}{\bibinfo{person}{Paul Covington}, \bibinfo{person}{Jay Adams}, {and} \bibinfo{person}{Emre Sargin}.} \bibinfo{year}{2016}\natexlab{}.
\newblock \showarticletitle{Deep Neural Networks for YouTube Recommendations}. In \bibinfo{booktitle}{\emph{Proceedings of the 10th {ACM} Conference on Recommender Systems, Boston, MA, USA, September 15-19, 2016}}, \bibfield{editor}{\bibinfo{person}{Shilad Sen}, \bibinfo{person}{Werner Geyer}, \bibinfo{person}{Jill Freyne}, {and} \bibinfo{person}{Pablo Castells}} (Eds.). \bibinfo{publisher}{{ACM}}, \bibinfo{pages}{191--198}.
\newblock
\urldef\tempurl%
\url{https://doi.org/10.1145/2959100.2959190}
\showDOI{\tempurl}


\bibitem[He and Chua(2017)]%
        {cf_1}
\bibfield{author}{\bibinfo{person}{Xiangnan He} {and} \bibinfo{person}{Tat{-}Seng Chua}.} \bibinfo{year}{2017}\natexlab{}.
\newblock \showarticletitle{Neural Factorization Machines for Sparse Predictive Analytics}. In \bibinfo{booktitle}{\emph{Proceedings of the 40th International {ACM} {SIGIR} Conference on Research and Development in Information Retrieval, Shinjuku, Tokyo, Japan, August 7-11, 2017}}, \bibfield{editor}{\bibinfo{person}{Noriko Kando}, \bibinfo{person}{Tetsuya Sakai}, \bibinfo{person}{Hideo Joho}, \bibinfo{person}{Hang Li}, \bibinfo{person}{Arjen~P. de~Vries}, {and} \bibinfo{person}{Ryen~W. White}} (Eds.). \bibinfo{publisher}{{ACM}}, \bibinfo{pages}{355--364}.
\newblock
\urldef\tempurl%
\url{https://doi.org/10.1145/3077136.3080777}
\showDOI{\tempurl}


\bibitem[Huang et~al\mbox{.}(2016)]%
        {cf_2}
\bibfield{author}{\bibinfo{person}{Yanxiang Huang}, \bibinfo{person}{Bin Cui}, \bibinfo{person}{Jie Jiang}, \bibinfo{person}{Kunqian Hong}, \bibinfo{person}{Wenyu Zhang}, {and} \bibinfo{person}{Yiran Xie}.} \bibinfo{year}{2016}\natexlab{}.
\newblock \showarticletitle{Real-time video recommendation exploration}. In \bibinfo{booktitle}{\emph{Proceedings of the 2016 international conference on management of data}}. \bibinfo{pages}{35--46}.
\newblock


\bibitem[Huh et~al\mbox{.}(2024)]%
        {platonic}
\bibfield{author}{\bibinfo{person}{Minyoung Huh}, \bibinfo{person}{Brian Cheung}, \bibinfo{person}{Tongzhou Wang}, {and} \bibinfo{person}{Phillip Isola}.} \bibinfo{year}{2024}\natexlab{}.
\newblock \showarticletitle{The Platonic Representation Hypothesis}.
\newblock \bibinfo{journal}{\emph{CoRR}}  \bibinfo{volume}{abs/2405.07987} (\bibinfo{year}{2024}).
\newblock
\urldef\tempurl%
\url{https://doi.org/10.48550/ARXIV.2405.07987}
\showDOI{\tempurl}
\showeprint[arXiv]{2405.07987}


\bibitem[Jing et~al\mbox{.}(2024)]%
        {micro_MARNET}
\bibfield{author}{\bibinfo{person}{Peiguang Jing}, \bibinfo{person}{Xianyi Liu}, \bibinfo{person}{Lijuan Zhang}, \bibinfo{person}{Yun Li}, \bibinfo{person}{Yu Liu}, {and} \bibinfo{person}{Yuting Su}.} \bibinfo{year}{2024}\natexlab{}.
\newblock \showarticletitle{Multimodal Attentive Representation Learning for Micro-video Multi-label Classification}.
\newblock \bibinfo{journal}{\emph{{ACM} Trans. Multim. Comput. Commun. Appl.}} \bibinfo{volume}{20}, \bibinfo{number}{6} (\bibinfo{year}{2024}), \bibinfo{pages}{182:1--182:23}.
\newblock
\urldef\tempurl%
\url{https://doi.org/10.1145/3643888}
\showDOI{\tempurl}


\bibitem[Liu et~al\mbox{.}(2024)]%
        {mmr_trans}
\bibfield{author}{\bibinfo{person}{Han Liu}, \bibinfo{person}{Yinwei Wei}, \bibinfo{person}{Xuemeng Song}, \bibinfo{person}{Weili Guan}, \bibinfo{person}{Yuan{-}Fang Li}, {and} \bibinfo{person}{Liqiang Nie}.} \bibinfo{year}{2024}\natexlab{}.
\newblock \showarticletitle{MMGRec: Multimodal Generative Recommendation with Transformer Model}.
\newblock \bibinfo{journal}{\emph{CoRR}}  \bibinfo{volume}{abs/2404.16555} (\bibinfo{year}{2024}).
\newblock
\urldef\tempurl%
\url{https://doi.org/10.48550/ARXIV.2404.16555}
\showDOI{\tempurl}
\showeprint[arXiv]{2404.16555}


\bibitem[Liu et~al\mbox{.}(2020)]%
        {video_gnn}
\bibfield{author}{\bibinfo{person}{Qi Liu}, \bibinfo{person}{Ruobing Xie}, \bibinfo{person}{Lei Chen}, \bibinfo{person}{Shukai Liu}, \bibinfo{person}{Ke Tu}, \bibinfo{person}{Peng Cui}, \bibinfo{person}{Bo Zhang}, {and} \bibinfo{person}{Leyu Lin}.} \bibinfo{year}{2020}\natexlab{}.
\newblock \showarticletitle{Graph neural network for tag ranking in tag-enhanced video recommendation}. In \bibinfo{booktitle}{\emph{Proceedings of the 29th ACM international conference on information \& knowledge management}}. \bibinfo{pages}{2613--2620}.
\newblock


\bibitem[Pi et~al\mbox{.}(2020)]%
        {SIM}
\bibfield{author}{\bibinfo{person}{Qi Pi}, \bibinfo{person}{Guorui Zhou}, \bibinfo{person}{Yujing Zhang}, \bibinfo{person}{Zhe Wang}, \bibinfo{person}{Lejian Ren}, \bibinfo{person}{Ying Fan}, \bibinfo{person}{Xiaoqiang Zhu}, {and} \bibinfo{person}{Kun Gai}.} \bibinfo{year}{2020}\natexlab{}.
\newblock \showarticletitle{Search-based User Interest Modeling with Lifelong Sequential Behavior Data for Click-Through Rate Prediction}. In \bibinfo{booktitle}{\emph{{CIKM} '20: The 29th {ACM} International Conference on Information and Knowledge Management, Virtual Event, Ireland, October 19-23, 2020}}, \bibfield{editor}{\bibinfo{person}{Mathieu d'Aquin}, \bibinfo{person}{Stefan Dietze}, \bibinfo{person}{Claudia Hauff}, \bibinfo{person}{Edward Curry}, {and} \bibinfo{person}{Philippe Cudr{\'{e}}{-}Mauroux}} (Eds.). \bibinfo{publisher}{{ACM}}, \bibinfo{pages}{2685--2692}.
\newblock
\urldef\tempurl%
\url{https://doi.org/10.1145/3340531.3412744}
\showDOI{\tempurl}


\bibitem[Quan et~al\mbox{.}(2024)]%
        {micro_len}
\bibfield{author}{\bibinfo{person}{Yuhan Quan}, \bibinfo{person}{Jingtao Ding}, \bibinfo{person}{Chen Gao}, \bibinfo{person}{Nian Li}, \bibinfo{person}{Lingling Yi}, \bibinfo{person}{Depeng Jin}, {and} \bibinfo{person}{Yong Li}.} \bibinfo{year}{2024}\natexlab{}.
\newblock \showarticletitle{Alleviating Video-length Effect for Micro-video Recommendation}.
\newblock \bibinfo{journal}{\emph{{ACM} Trans. Inf. Syst.}} \bibinfo{volume}{42}, \bibinfo{number}{2} (\bibinfo{year}{2024}), \bibinfo{pages}{44:1--44:24}.
\newblock
\urldef\tempurl%
\url{https://doi.org/10.1145/3617826}
\showDOI{\tempurl}


\bibitem[Shang et~al\mbox{.}(2023)]%
        {micro_fine}
\bibfield{author}{\bibinfo{person}{Yu Shang}, \bibinfo{person}{Chen Gao}, \bibinfo{person}{Jiansheng Chen}, \bibinfo{person}{Depeng Jin}, \bibinfo{person}{Meng Wang}, {and} \bibinfo{person}{Yong Li}.} \bibinfo{year}{2023}\natexlab{}.
\newblock \showarticletitle{Learning Fine-grained User Interests for Micro-video Recommendation}. In \bibinfo{booktitle}{\emph{Proceedings of the 46th International {ACM} {SIGIR} Conference on Research and Development in Information Retrieval, {SIGIR} 2023, Taipei, Taiwan, July 23-27, 2023}}, \bibfield{editor}{\bibinfo{person}{Hsin{-}Hsi Chen}, \bibinfo{person}{Wei{-}Jou~(Edward) Duh}, \bibinfo{person}{Hen{-}Hsen Huang}, \bibinfo{person}{Makoto~P. Kato}, \bibinfo{person}{Josiane Mothe}, {and} \bibinfo{person}{Barbara Poblete}} (Eds.). \bibinfo{publisher}{{ACM}}, \bibinfo{pages}{433--442}.
\newblock
\urldef\tempurl%
\url{https://doi.org/10.1145/3539618.3591713}
\showDOI{\tempurl}


\bibitem[Wang et~al\mbox{.}(2019b)]%
        {cf_3}
\bibfield{author}{\bibinfo{person}{Meirui Wang}, \bibinfo{person}{Pengjie Ren}, \bibinfo{person}{Lei Mei}, \bibinfo{person}{Zhumin Chen}, \bibinfo{person}{Jun Ma}, {and} \bibinfo{person}{Maarten de Rijke}.} \bibinfo{year}{2019}\natexlab{b}.
\newblock \showarticletitle{A Collaborative Session-based Recommendation Approach with Parallel Memory Modules}. In \bibinfo{booktitle}{\emph{Proceedings of the 42nd International {ACM} {SIGIR} Conference on Research and Development in Information Retrieval, {SIGIR} 2019, Paris, France, July 21-25, 2019}}, \bibfield{editor}{\bibinfo{person}{Benjamin Piwowarski}, \bibinfo{person}{Max Chevalier}, \bibinfo{person}{{\'{E}}ric Gaussier}, \bibinfo{person}{Yoelle Maarek}, \bibinfo{person}{Jian{-}Yun Nie}, {and} \bibinfo{person}{Falk Scholer}} (Eds.). \bibinfo{publisher}{{ACM}}, \bibinfo{pages}{345--354}.
\newblock
\urldef\tempurl%
\url{https://doi.org/10.1145/3331184.3331210}
\showDOI{\tempurl}


\bibitem[Wang et~al\mbox{.}(2019a)]%
        {cf_4}
\bibfield{author}{\bibinfo{person}{Pengfei Wang}, \bibinfo{person}{Hanxiong Chen}, \bibinfo{person}{Yadong Zhu}, \bibinfo{person}{Huawei Shen}, {and} \bibinfo{person}{Yongfeng Zhang}.} \bibinfo{year}{2019}\natexlab{a}.
\newblock \showarticletitle{Unified Collaborative Filtering over Graph Embeddings}. In \bibinfo{booktitle}{\emph{Proceedings of the 42nd International {ACM} {SIGIR} Conference on Research and Development in Information Retrieval, {SIGIR} 2019, Paris, France, July 21-25, 2019}}, \bibfield{editor}{\bibinfo{person}{Benjamin Piwowarski}, \bibinfo{person}{Max Chevalier}, \bibinfo{person}{{\'{E}}ric Gaussier}, \bibinfo{person}{Yoelle Maarek}, \bibinfo{person}{Jian{-}Yun Nie}, {and} \bibinfo{person}{Falk Scholer}} (Eds.). \bibinfo{publisher}{{ACM}}, \bibinfo{pages}{155--164}.
\newblock
\urldef\tempurl%
\url{https://doi.org/10.1145/3331184.3331224}
\showDOI{\tempurl}


\bibitem[Wei et~al\mbox{.}(2020)]%
        {content_3}
\bibfield{author}{\bibinfo{person}{Yinwei Wei}, \bibinfo{person}{Xiang Wang}, \bibinfo{person}{Weili Guan}, \bibinfo{person}{Liqiang Nie}, \bibinfo{person}{Zhouchen Lin}, {and} \bibinfo{person}{Baoquan Chen}.} \bibinfo{year}{2020}\natexlab{}.
\newblock \showarticletitle{Neural Multimodal Cooperative Learning Toward Micro-Video Understanding}.
\newblock \bibinfo{journal}{\emph{{IEEE} Trans. Image Process.}}  \bibinfo{volume}{29} (\bibinfo{year}{2020}), \bibinfo{pages}{1--14}.
\newblock
\urldef\tempurl%
\url{https://doi.org/10.1109/TIP.2019.2923608}
\showDOI{\tempurl}


\bibitem[Zhan et~al\mbox{.}(2022)]%
        {D2Q}
\bibfield{author}{\bibinfo{person}{Ruohan Zhan}, \bibinfo{person}{Changhua Pei}, \bibinfo{person}{Qiang Su}, \bibinfo{person}{Jianfeng Wen}, \bibinfo{person}{Xueliang Wang}, \bibinfo{person}{Guanyu Mu}, \bibinfo{person}{Dong Zheng}, \bibinfo{person}{Peng Jiang}, {and} \bibinfo{person}{Kun Gai}.} \bibinfo{year}{2022}\natexlab{}.
\newblock \showarticletitle{Deconfounding Duration Bias in Watch-time Prediction for Video Recommendation}. In \bibinfo{booktitle}{\emph{{KDD} '22: The 28th {ACM} {SIGKDD} Conference on Knowledge Discovery and Data Mining, Washington, DC, USA, August 14 - 18, 2022}}, \bibfield{editor}{\bibinfo{person}{Aidong Zhang} {and} \bibinfo{person}{Huzefa Rangwala}} (Eds.). \bibinfo{publisher}{{ACM}}, \bibinfo{pages}{4472--4481}.
\newblock
\urldef\tempurl%
\url{https://doi.org/10.1145/3534678.3539092}
\showDOI{\tempurl}


\bibitem[Zhang et~al\mbox{.}(2023)]%
        {mmr_con}
\bibfield{author}{\bibinfo{person}{Jinghao Zhang}, \bibinfo{person}{Yanqiao Zhu}, \bibinfo{person}{Qiang Liu}, \bibinfo{person}{Mengqi Zhang}, \bibinfo{person}{Shu Wu}, {and} \bibinfo{person}{Liang Wang}.} \bibinfo{year}{2023}\natexlab{}.
\newblock \showarticletitle{Latent Structure Mining With Contrastive Modality Fusion for Multimedia Recommendation}.
\newblock \bibinfo{journal}{\emph{{IEEE} Trans. Knowl. Data Eng.}} \bibinfo{volume}{35}, \bibinfo{number}{9} (\bibinfo{year}{2023}), \bibinfo{pages}{9154--9167}.
\newblock
\urldef\tempurl%
\url{https://doi.org/10.1109/TKDE.2022.3221949}
\showDOI{\tempurl}


\bibitem[Zhao et~al\mbox{.}(2023)]%
        {D2Co}
\bibfield{author}{\bibinfo{person}{Haiyuan Zhao}, \bibinfo{person}{Lei Zhang}, \bibinfo{person}{Jun Xu}, \bibinfo{person}{Guohao Cai}, \bibinfo{person}{Zhenhua Dong}, {and} \bibinfo{person}{Ji{-}Rong Wen}.} \bibinfo{year}{2023}\natexlab{}.
\newblock \showarticletitle{Uncovering User Interest from Biased and Noised Watch Time in Video Recommendation}. In \bibinfo{booktitle}{\emph{Proceedings of the 17th {ACM} Conference on Recommender Systems, RecSys 2023, Singapore, Singapore, September 18-22, 2023}}, \bibfield{editor}{\bibinfo{person}{Jie Zhang}, \bibinfo{person}{Li~Chen}, \bibinfo{person}{Shlomo Berkovsky}, \bibinfo{person}{Min Zhang}, \bibinfo{person}{Tommaso~Di Noia}, \bibinfo{person}{Justin Basilico}, \bibinfo{person}{Luiz Pizzato}, {and} \bibinfo{person}{Yang Song}} (Eds.). \bibinfo{publisher}{{ACM}}, \bibinfo{pages}{528--539}.
\newblock
\urldef\tempurl%
\url{https://doi.org/10.1145/3604915.3608797}
\showDOI{\tempurl}


\bibitem[Zheng et~al\mbox{.}(2022)]%
        {WTG}
\bibfield{author}{\bibinfo{person}{Yu Zheng}, \bibinfo{person}{Chen Gao}, \bibinfo{person}{Jingtao Ding}, \bibinfo{person}{Lingling Yi}, \bibinfo{person}{Depeng Jin}, \bibinfo{person}{Yong Li}, {and} \bibinfo{person}{Meng Wang}.} \bibinfo{year}{2022}\natexlab{}.
\newblock \showarticletitle{{DVR:} Micro-Video Recommendation Optimizing Watch-Time-Gain under Duration Bias}. In \bibinfo{booktitle}{\emph{{MM} '22: The 30th {ACM} International Conference on Multimedia, Lisboa, Portugal, October 10 - 14, 2022}}, \bibfield{editor}{\bibinfo{person}{Jo{\~{a}}o Magalh{\~{a}}es}, \bibinfo{person}{Alberto~Del Bimbo}, \bibinfo{person}{Shin'ichi Satoh}, \bibinfo{person}{Nicu Sebe}, \bibinfo{person}{Xavier Alameda{-}Pineda}, \bibinfo{person}{Qin Jin}, \bibinfo{person}{Vincent Oria}, {and} \bibinfo{person}{Laura Toni}} (Eds.). \bibinfo{publisher}{{ACM}}, \bibinfo{pages}{334--345}.
\newblock
\urldef\tempurl%
\url{https://doi.org/10.1145/3503161.3548428}
\showDOI{\tempurl}


\bibitem[Zhong et~al\mbox{.}(2024)]%
        {mirror}
\bibfield{author}{\bibinfo{person}{Shanshan Zhong}, \bibinfo{person}{Zhongzhan Huang}, \bibinfo{person}{Daifeng Li}, \bibinfo{person}{Wushao Wen}, \bibinfo{person}{Jinghui Qin}, {and} \bibinfo{person}{Liang Lin}.} \bibinfo{year}{2024}\natexlab{}.
\newblock \showarticletitle{Mirror Gradient: Towards Robust Multimodal Recommender Systems via Exploring Flat Local Minima}. In \bibinfo{booktitle}{\emph{Proceedings of the {ACM} on Web Conference 2024, {WWW} 2024, Singapore, May 13-17, 2024}}, \bibfield{editor}{\bibinfo{person}{Tat{-}Seng Chua}, \bibinfo{person}{Chong{-}Wah Ngo}, \bibinfo{person}{Ravi Kumar}, \bibinfo{person}{Hady~W. Lauw}, {and} \bibinfo{person}{Roy~Ka{-}Wei Lee}} (Eds.). \bibinfo{publisher}{{ACM}}, \bibinfo{pages}{3700--3711}.
\newblock
\urldef\tempurl%
\url{https://doi.org/10.1145/3589334.3645553}
\showDOI{\tempurl}


\bibitem[Zhou et~al\mbox{.}(2023c)]%
        {dragon}
\bibfield{author}{\bibinfo{person}{Hongyu Zhou}, \bibinfo{person}{Xin Zhou}, \bibinfo{person}{Lingzi Zhang}, {and} \bibinfo{person}{Zhiqi Shen}.} \bibinfo{year}{2023}\natexlab{c}.
\newblock \showarticletitle{Enhancing dyadic relations with homogeneous graphs for multimodal recommendation}.
\newblock In \bibinfo{booktitle}{\emph{ECAI 2023}}. \bibinfo{publisher}{IOS Press}, \bibinfo{pages}{3123--3130}.
\newblock


\bibitem[Zhou and Miao(2024)]%
        {mmr_vae}
\bibfield{author}{\bibinfo{person}{Xin Zhou} {and} \bibinfo{person}{Chunyan Miao}.} \bibinfo{year}{2024}\natexlab{}.
\newblock \showarticletitle{Disentangled Graph Variational Auto-Encoder for Multimodal Recommendation With Interpretability}.
\newblock \bibinfo{journal}{\emph{{IEEE} Trans. Multim.}}  \bibinfo{volume}{26} (\bibinfo{year}{2024}), \bibinfo{pages}{7543--7554}.
\newblock
\urldef\tempurl%
\url{https://doi.org/10.1109/TMM.2024.3369875}
\showDOI{\tempurl}


\bibitem[Zhou and Shen(2023)]%
        {freedom}
\bibfield{author}{\bibinfo{person}{Xin Zhou} {and} \bibinfo{person}{Zhiqi Shen}.} \bibinfo{year}{2023}\natexlab{}.
\newblock \showarticletitle{A tale of two graphs: Freezing and denoising graph structures for multimodal recommendation}. In \bibinfo{booktitle}{\emph{Proceedings of the 31st ACM International Conference on Multimedia}}. \bibinfo{pages}{935--943}.
\newblock


\bibitem[Zhou et~al\mbox{.}(2023b)]%
        {BM3}
\bibfield{author}{\bibinfo{person}{Xin Zhou}, \bibinfo{person}{Hongyu Zhou}, \bibinfo{person}{Yong Liu}, \bibinfo{person}{Zhiwei Zeng}, \bibinfo{person}{Chunyan Miao}, \bibinfo{person}{Pengwei Wang}, \bibinfo{person}{Yuan You}, {and} \bibinfo{person}{Feijun Jiang}.} \bibinfo{year}{2023}\natexlab{b}.
\newblock \showarticletitle{Bootstrap latent representations for multi-modal recommendation}. In \bibinfo{booktitle}{\emph{Proceedings of the ACM Web Conference 2023}}. \bibinfo{pages}{845--854}.
\newblock


\bibitem[Zhou et~al\mbox{.}(2023a)]%
        {mmr_att}
\bibfield{author}{\bibinfo{person}{Yan Zhou}, \bibinfo{person}{Jie Guo}, \bibinfo{person}{Hao Sun}, \bibinfo{person}{Bin Song}, {and} \bibinfo{person}{Fei~Richard Yu}.} \bibinfo{year}{2023}\natexlab{a}.
\newblock \showarticletitle{Attention-guided Multi-step Fusion: {A} Hierarchical Fusion Network for Multimodal Recommendation}. In \bibinfo{booktitle}{\emph{Proceedings of the 46th International {ACM} {SIGIR} Conference on Research and Development in Information Retrieval, {SIGIR} 2023, Taipei, Taiwan, July 23-27, 2023}}, \bibfield{editor}{\bibinfo{person}{Hsin{-}Hsi Chen}, \bibinfo{person}{Wei{-}Jou~(Edward) Duh}, \bibinfo{person}{Hen{-}Hsen Huang}, \bibinfo{person}{Makoto~P. Kato}, \bibinfo{person}{Josiane Mothe}, {and} \bibinfo{person}{Barbara Poblete}} (Eds.). \bibinfo{publisher}{{ACM}}, \bibinfo{pages}{1816--1820}.
\newblock
\urldef\tempurl%
\url{https://doi.org/10.1145/3539618.3591950}
\showDOI{\tempurl}


\end{thebibliography}

\appendix









\end{document}